\documentclass[useAMS]{gGAF2e}
\usepackage{graphicx,amsmath,url}
\usepackage{bm}
\graphicspath{{./fig/}{./png/}}
\newcommand{\parr}[2]{\frac{\partial #1}{\partial #2}}

\newcommand{\hhh}{h}

\newcommand{\hhM}{\overline{h}_{\rm T}}

\newcommand {\AAAM}{\overline{\AAA}}
\newcommand {\BBM}{\overline{\BB}}
\newcommand {\JJM}{\overline{\JJ}}

\newcommand {\EMFM}{\overline{\EMF}}

\newcommand{\aaa}{\mbox{\boldmath $a$}{}}{}

\newcommand{\hhm}{\overline{h}_{\text{m}}}
\newcommand{\hhf}{\overline{h}_{\text{f}}}

\newcommand{\EQ}{\begin{equation}}
\newcommand{\EN}{\end{equation}}
\newcommand{\EQA}{\begin{eqnarray}}
\newcommand{\ENA}{\end{eqnarray}}

\newcommand{\Eq}[1]{Equation~(\ref{#1})}
\newcommand{\Eqs}[2]{Equations~(\ref{#1}) and~(\ref{#2})}

\newcommand{\Sec}[1]{Section~\ref{#1}}

\newcommand{\Fig}[1]{Figure~\ref{#1}}

\newcommand{\Tab}[1]{Table~\ref{#1}}

\newcommand{\bra}[1]{\langle #1\rangle}
\newcommand{\bbra}[1]{\left\langle #1\right\rangle}

{}
{}
\newcommand{\meanFFFF}{\overline{\mbox{\boldmath ${\cal F}$}}{}}{}

{}
{}
\newcommand{\meanEMF}{\overline{\mbox{\boldmath ${\cal E}$}}{}}{}
{}
{}
{}
{}
{}
\newcommand{\meanAA}{\overline{\mbox{\boldmath $A$}}{}}{}
\newcommand{\meanBB}{\overline{\mbox{\boldmath $B$}}{}}{}
\newcommand{\meanEE}{\overline{\mbox{\boldmath $E$}}{}}{}
{}
{}
{}
{}
{}
{}
{}
\newcommand{\meanJJ}{\overline{\mbox{\boldmath $J$}}{}}{}

{}
{}
{}

\newcommand{\meanh}{\overline{h}}

{}

{}
{}

\newcommand{\alphaK}{\alpha_{\rm K}}
\newcommand{\alphaM}{\alpha_{\rm M}}
\newcommand{\zzz}{\hat{\mbox{\boldmath $z$}} {}}

\newcommand{\uu}{\mbox{\boldmath $u$} {}}

\def\bb{\bm{b}}
\newcommand{\BB}{\mbox{\boldmath $B$} {}}
\newcommand{\EE}{\mbox{\boldmath $E$} {}}
\newcommand{\jj}{\mbox{\boldmath $j$} {}}
\newcommand{\JJ}{\mbox{\boldmath $J$} {}}

\newcommand{\AAA}{\mbox{\boldmath $A$} {}}
\newcommand{\aaaa}{\mbox{\boldmath $a$} {}}
\newcommand{\ee}{\mbox{\boldmath $e$} {}}

\newcommand{\ff}{\mbox{\boldmath $f$} {}}

\newcommand{\FFF}{\mbox{\boldmath ${\cal F}$} {}}

\newcommand{\nab}{\mbox{\boldmath $\nabla$} {}}

\newcommand{\EMF}{\mbox{\boldmath ${\cal E}$} {}}

\newcommand{\dd}{{\rm d} {}}

\def\la{\mathrel{\mathchoice {\vcenter{\offinterlineskip\halign{\hfil
$\displaystyle##$\hfil\cr<\cr\sim\cr}}}
{\vcenter{\offinterlineskip\halign{\hfil$\textstyle##$\hfil\cr<\cr\sim\cr}}}
{\vcenter{\offinterlineskip\halign{\hfil$\scriptstyle##$\hfil\cr<\cr\sim\cr}}}
{\vcenter{\offinterlineskip\halign{\hfil$\scriptscriptstyle##$\hfil\cr<\cr\sim\cr}}}}}

\def\Rm{R_{\rm m}}

\def\kf{k_{\rm f}}

\def\urms{u_{\rm rms}}

\def\etat{\eta_{\rm t}}
\def\etatz{\eta_{\rm t0}}

\def\Beq{B_{\rm eq}}
\newcommand{\yan}[5]{, ``#5,'' {\em Astron.\ Nachr.\ }{\bf #2}, #3-#4 (#1).}

\newcommand{\yana}[5]{, ``#5,'' {\em Astron.\ Astrophys.\ }{\bf #2}, #3-#4 (#1).}

\newcommand{\ymn}[5]{, ``#5,'' {\em Monthly Notices Roy.\ Astron.\ Soc.\ }{\bf #2}, #3-#4 (#1).}

\newcommand{\yjfm}[5]{, ``#5,'' {\em J.\ Fluid Mech.\ }{\bf #2}, #3-#4 (#1).}

\newcommand{\ypre}[5]{, ``#5,'' {\em Phys.\ Rev.\ E }{\bf #2}, #3-#4 (#1).}
\newcommand{\ypreN}[4]{, ``#4,'' {\em Phys.\ Rev.\ }{\bf #2}, #3 (#1).}
\newcommand{\yprl}[5]{, ``#5,'' {\em Phys.\ Rev.\ Letters }{\bf #2}, #3-#4 (#1).}
\newcommand{\yprlN}[4]{, ``#4,'' {\em Phys.\ Rev.\ Letters }{\bf #2}, #3 (#1).}

\newcommand{\yapj}[5]{, ``#5,'' {\em Astrophys.\ J.\ }{\bf #2}, #3-#4 (#1).}

\newcommand{\yapjl}[5]{, ``#5,'' {\em Astrophys.\ J.\ Letters }{\bf #2}, #3-#4 (#1).}

\newcommand{\ygafd}[5]{, ``#5,'' {\em Geophys.\ Astrophys.\ Fluid Dynam. }{\bf #2}, #3-#4 (#1).}

\newcommand{\yjour}[6]{, ``#6,'' {\em #2} {\bf #3}, #4-#5 (#1).}

\newcommand{\yproc}[7]{, ``#4,'' In {\em #5} (ed.\ #6), pp.\ #2-#3.\ #7 (#1).}
\newcommand{\ybook}[3]{ {\em #2}.\ #3 (#1).}

\newcommand{\sana}[2]{ ~#1~ ``#2,'' {\em Astron.\ Astrophys.\ } (submitted).}

\begin{document}
\doi{}
\issn{} \issnp{} \jvol{00} \jnum{00} \jyear{2010}

\markboth{A. Hubbard and A. Brandenburg}{Magnetic helicity fluxes with a halo}

\title{Magnetic helicity fluxes in an $\alpha^2$ dynamo embedded in a halo}
\author{Alexander Hubbard${\dag}$ and Axel Brandenburg${\dag\ddag}$\\
\vspace{6pt}
${\dag}$NORDITA, AlbaNova University Center, Roslagstullsbacken 23,
SE 10691 Stockholm, Sweden\\
${\ddag}$Department of Astronomy,
Stockholm University, SE 10691 Stockholm, Sweden}

\date{\today,~ $ $Revision: 1.76 $ $}

\maketitle

\begin{abstract}
We present the results of simulations of forced turbulence
in a slab where the mean kinetic helicity has a maximum
near the mid-plane, generating gradients of magnetic helicity of both
large and small-scale fields.
We also study systems that have poorly conducting buffer zones
away from the midplane in order to assess the effects of boundaries. 
The dynamical $\alpha$ quenching phenomenology requires that the magnetic helicity in the
small-scale fields approaches a nearly static, gauge independent state.
To stress-test this steady state condition we choose a system with a uniform
sign of kinetic helicity, so
that the total magnetic helicity can reach a steady state value only through fluxes
through the boundary, which are themselves suppressed by the velocity boundary conditions.
Even with such a set up, the small-scale magnetic helicity
is found to reach a steady state.
In agreement with earlier work, the magnetic helicity fluxes of small-scale
fields are found to be turbulently diffusive.
By comparing results with and without halos,
we show that artificial constraints on magnetic helicity at the boundary do not
have a significant impact on the evolution
of the magnetic helicity, except that ``softer" (halo) boundary conditions
give a lower energy of the saturated mean magnetic field.
\end{abstract}


\section{Introduction}

Stars with outer convection zones tend to possess magnetic fields that
display spatio-temporal order with variations that are often cyclic and,
in the case of the Sun, antisymmetric with respect to the equatorial plane.
Simulations now begin to reproduce much of this behavior
\citep[see, e.g.,][]{Brown10,Kapyla10,GCS10}.
A useful tool for understanding the outcomes of such models is
mean-field dynamo theory.
A central ingredient of this theory is the $\alpha$ effect.
This effect quantifies a component of the mean electromotive force
that is proportional to the mean magnetic field \citep{Mof78,KR80}.

Mean-field theory gives meaningful predictions when to expect cyclic or
steady solutions, and what the symmetry properties with respect to the
equator are \citep{B98}.
Even in the nonlinear regime, the simple concept of $\alpha$ quenching,
which reduces $\alpha$ locally via an algebraic function of the mean
magnetic field, tends to give plausible results.
However, under some circumstances, it becomes quite clear that this
simple-minded approach must be wrong.
Such a special case is that of a triply-periodic domain.
Astrophysically speaking, such a model is quite unrealistic, but it
is often employed in numerical simulations.
It was also employed as the primary tool to compute $\alpha$ quenching
from simulations \citep{CH96}.
These simulations suggest that $\alpha$ quenching would set in once
the mean field becomes comparable to a small fraction ($\Rm^{-1/2}$, where
$\Rm$ is the magnetic Reynolds number) times the equipartition value.
If this were true also for astrophysical bodies such as the Sun,
the $\alpha$ effect could not be invoked for understanding the
dynamics of the Sun's magnetic field.

Later it became clear that there are counter examples to the simple idea
that $\alpha$ is quenched just depending on the local field strength.
Surprisingly, simulations later suggested that even in a triply-periodic
domain a large-scale magnetic field can be generated that can
even exceed the equipartition value \citep{B01}.
However, it would take a resistive time-scale to reach these field
strengths, so there was still a problem.
Around the same time, the idea emerged that open boundaries might help
\citep{BF00a,BF00b,KMRS00,KMRS02}.
This is connected with the fact that an $\alpha$ effect dynamo produces
magnetic helicity of opposite sign at large and small scales
\citep{See96,Ji99}.
The magnetic helicity at small scales is an unwanted by-product that
can feed back adversely on the dynamo.
The resistively slow saturation phase in periodic-box simulations can
then be understood in terms of the time it takes to dissipate this
small-scale magnetic helicity.
It is indeed a particular property of triply-periodic domains that magnetic
helicity is strictly conserved at large magnetic Reynolds numbers.
A possible remedy might then be to consider open domains that allow
magnetic helicity fluxes.

The first simulations with open domains were not encouraging.
While it was possible to reach saturation more quickly, the field was
found to level off at a value that becomes progressively smaller at
larger magnetic Reynolds numbers \citep{BD01,BS05}.
A possible problem with these simulations might be the absence of
magnetic helicity fluxes within the domain.
Indeed, \cite{BD01} considered a kinetic helicity distribution
that was approximately uniform across the domain, so there were
no gradients except in the immediate proximity of boundaries,
where boundary conditions on the velocity prevent turbulent diffusion.
The situation improved dramatically when simulations with shear
were considered \citep{B05,Kapyla08,HP09}.
Shear provides not only an additional induction effect for
the dynamo, but it might also lead to an additional source of
magnetic helicity flux within the domain \citep{VC01,SB04,SB06}.
More recently it turned out that, even without shear,
diffusion down the gradient of small-scale magnetic helicity
could, at least in principle, help avoid vanishingly small
saturation levels of the mean magnetic field when the magnetic
Reynolds number becomes large \citep{BCC09,Mitra10}.

An important goal of the present paper it to revisit this issue using
direct simulations of turbulent dynamos without shear, and even with
the same sign of magnetic helicity everywhere, but with a spatial
modulation of the helicity within the domain.
In other words, the level of turbulence is maintained at a high
level throughout the domain, but the amount of swirl diminishes toward
the boundaries.
In most of the simulations we include a turbulent halo outside the
dynamo domain where the Ohmic resistivity is enhanced.
This might be important as several simple boundary conditions such as
pseudo-vacuum (or vertical field) conditions fix the value
of the magnetic helicity artificially,
and if fluid motions through the boundary are prohibited, turbulent transport
there is impossible.

Our simulations also allow us to make contact with nonlinear mean-field phenomenology
where the evolution of the small-scale magnetic helicity is taken into account.
This leads then to an evolution equation for an additional contribution to the
$\alpha$ effect, $\alphaM$.
This approach is referred to as dynamical $\alpha$ quenching.
In the present paper we will also attempt to assess the
validity of some of the corner stones of dynamical $\alpha$ quenching.
Firstly, there
is the magnetic $\alpha$ of \cite{PFL}, where the fluctuating magnetic field
generates an $\alphaM$ that is proportional to the current helicity of
the fluctuating field.
This $\alphaM$ counteracts the kinetic $\alpha$, and so saturates the dynamo.
Secondly there is magnetic helicity conservation which notes that the total magnetic
helicity is nearly conserved under common conditions,
and so the helicity in the fluctuating field can be related to the helicity in the large-scale field.
Finally, there is the assumption that the mean current helicity of the fluctuating
field is proportional to the mean magnetic helicity in the fluctuating field.

As noted above, a problematic prediction of dynamical $\alpha$ quenching is that rapid
(exponential) growth of mean magnetic fields will be halted below equipartition
with the turbulent energy.  The export of small-scale helicity could provide a release
from this constraint but will likely occur side-by-side with export of the mean field.
The interplay between these effects can smother the dynamo
even in the presence of small-scale helicity transport.  Treatment of large-scale
helicity transport proves significantly more complicated than that of the small-scale
helicity, but we will draw some preliminary conclusions.

In \Sec{dyn} we discuss the dynamical $\alpha$ quenching phenomenology.
In \Sec{numerics}
we describe the numerical setup of the simulations whose results are analyzed in
\Sec{analysis}.  Mean-field models of the systems are discussed in \Sec{meanfield}
and we conclude in \Sec{conclusions}.

\section{Dynamical $\alpha$ quenching}
\label{dyn}

We wish to use a mean-field approach to the saturation behavior of dynamos.  In what follows
our averages will be denoted by overbars and the fluctuating terms will be
denoted by lower case symbols.  In the simulations we will be using planar $xy$ averaging
unless noted otherwise, so the mean magnetic vector potential is given by
\begin{eqnarray}
&&
\meanAA(z,t)=\int\int\AAA(x,y,z,t)\,\dd x\,\dd y/L_xL_y, \\
&& \AAA=\meanAA+\aaa,
\end{eqnarray}
so the mean magnetic field is $\meanBB=\nab\times\meanAA$ and
the mean current density is $\meanJJ=\nab\times\meanBB/\mu_0$,
where $\mu_0$ is the vacuum permeability.
In the following we adopt units in which $\mu_0=1$.
Throughout this paper we use the expressions `mean field' and `large-scale
field' synonymously.
Likewise, we refer to the `small-scale field' as the `fluctuating field'.

We will work in the Weyl gauge (zero electrostatic potential, i.e.\
$\partial \AAA/\partial t=\uu \times \BB-\eta \JJ$),
and assume that there is no mean velocity.
We adopt the magnetic $\alpha$ prescription of \cite{KR82}.
As such our mean-field theoretic equations are:
\begin{eqnarray}
&&\parr{\BBM}{t}=\nab \times( \EMFM-\eta \JJM), \\
&&\EMFM =\alpha \BBM-\etat \JJM, \label{EMF}\\
&& \alpha=\alphaK+\alphaM,\\
&&\alphaM=\frac{\tau}{3}\frac{\overline{\jj \cdot \bb}}{\mu_0\rho}
\simeq k_{\rm f}^2\frac{\tau}{3}\frac{\overline{ \aaa \cdot \bb}}{\mu_0\rho}
\simeq k_{\rm f}^2 \frac{\etat}{\Beq^2}\overline{ \aaa \cdot \bb},
\label{alpM1}
\end{eqnarray}
where $\EMFM=\overline{\uu\times\bb}$ is the mean electromotive force,
$\alphaK$ is the kinetic $\alpha$ effect,
$\Beq^2 \equiv \rho \urms^2$ is a measure of the turbulent kinetic energy
and $\etat \equiv \tau\urms^2/3$ is the turbulent diffusivity.
The parameter  $\kf$ is the wavenumber of the energy carrying scale of the turbulence.
This $\alphaM$ is
taken to be the back-reaction component of $\alpha$ when it is split into kinetic
and magnetic components \citep{PFL}.

Magnetic helicity conservation can be seen from the time evolution equation
of the magnetic helicity density $\hhh_{\rm T} \equiv \AAA \cdot \BB$,
\EQ
\parr{\hhh_{\rm T}}{t}=-2 \eta \JJ \cdot \BB-\nab \cdot \FFF_{\rm T}, \label{dHdt1}
\EN
where $\FFF_{\rm T}$ is the magnetic helicity flux.
The subscript T refers to total field, which is composed of mean (m)
and fluctuating (f) fields.
In systems where the flux of magnetic helicity can be neglected
(such as spatially homogeneous systems), and when
the magnetic Reynolds number $\Rm$ is large enough
(and therefore $\eta$ small) the magnetic helicity will be nearly conserved.
We define the large-scale and small-scale helicities as
\begin{eqnarray}
&&\hhm \equiv \AAAM \cdot \BBM, \\
&&\hhf \equiv \overline{\aaa \cdot \bb}=\hhM-\hhm.
\end{eqnarray}
Averaging \Eq{dHdt1} we arrive at:
\begin{eqnarray}
&&\parr{\hhM}{t}=\parr{\hhm}{t}+\parr{\hhf}{t}=-2\eta \JJM \cdot \BBM -2\eta \overline{\jj \cdot \bb}
-\nab \cdot \meanFFFF_{\rm T}, \label{HM} \\
&&\parr{\hhm}{t}=2 \EMFM \cdot \BBM -2 \eta \JJM \cdot \BBM -
\nab \cdot \meanFFFF_{\rm m}, \label{Hm} \\
&&\parr{\hhf}{t}=-2\EMFM \cdot \BBM -2 \eta \overline{ \jj \cdot \bb} -
\nab \cdot \meanFFFF_{\rm f}. \label{Hf}
\end{eqnarray}
In the spirit of mean-field theory, we will scale the fluxes to gradients of mean quantities.
We consider here only a diffusive helicity flux of the small-scale fields,
\EQ
\meanFFFF_{\rm f} \sim -\kappa_{\rm f}\nab\hhf, 
\EN
while the flux of large-scale helicity will be discussed in greater detail in \Sec{analysis}.
In view of \Eq{alpM1}, the evolution of $\hhf$ is basically equivalent to the
evolution of $\alphaM$.

\section{Numerical setup}
\label{numerics}

In this paper we present both direct numerical simulations and
mean-field calculations.
In both cases we use the
{\sc Pencil Code}\footnote{\texttt{http://pencil-code.googlecode.com}},
which is a modular high-order code (sixth order in space and third-order
in time) for solving a large range of different partial differential
equations.

We consider models with and without a halo.
In both cases the horizontal extent of the domain is $L_x\times L_y$
with equal side lengths $L_x=L_y\equiv L$, with periodic boundaries.
In cases without a halo the vertical extent of the domain is $L_z=L$
while in cases with a halo we choose $L_z=2L$.
In the following we measure length in units of $k_1^{-1}$, where
$k_1=2\pi/L$ is the minimal horizontal wavenumber.
We will define our magnetic Reynolds number as
\EQ
\Rm \equiv \urms/\eta\kf,
\EN
where we assume that the wavenumber of the forcing is also
the wavenumber of the turbulence, and use a scale separation ratio $k_f/k_1=3$.

At the top and bottom, we impose stress-free velocity conditions
with $u_z=0=\partial u_x/\partial z =\partial u_y/\partial_z$.
  At the top and bottom we impose a ``vertical field" condition,
$A_z=\partial A_x /\partial z=\partial A_y /\partial z$.  This condition imposes $B_x=B_y=0$ and
hence $\AAA \cdot \BB=0$.  As this condition on magnetic helicity may be artificial, and the velocity
boundary condition constrains turbulent transport into the boundary, we include
buffer ``halos", such that the microscopic magnetic diffusivity is given by
\begin{eqnarray}
\eta=\begin{cases}
\eta_0 & -\pi \leq k_1 z \leq \pi \\
\eta_{\rm H} &|k_1 z|>\pi,
\end{cases}
\end{eqnarray}
where $\eta_{\rm H} \gg \eta_0$.
We include forced turbulence, with uniform amplitude and wavenumber $k_{\rm f}$,
but a relative helicity $\sigma=(\nab \times \ff \cdot \ff)/k_{\rm f} \ff^2$ of
\begin{eqnarray}
\sigma=\begin{cases}
\cos k_1z/2 & -\pi\leq  z \leq \pi \\
0 &|k_1 z|>\pi.
\end{cases}
\end{eqnarray}
For detailed about the implementation of a forcing function with
variable helicity we refer to the paper by \cite{HBD04}.

The above system is interesting from a dynamical $\alpha$ perspective
as it contains several contrasting elements.  Unlike \cite{Mitra10}, the forcing
helicity is all of one sign, and so we expect the magnetic and current helicities
to also be of one sign.  This implies that the production term in \Eq{dHdt1} be finite
even after volume averaging, and there may never be a final steady state for
the magnetic helicity.
Further, fluxes through the actual boundaries are reduced
through the velocity boundary conditions as well as the 
resistive destruction of the field in the halos.
It is not clear whether the
magnetic vector potential will even have a final steady solution.
An example of an unsteady magnetic helicity in an otherwise fully
steady dynamo was presented in Fig.~2 of \cite{BDS02}.

The main difference compared with earlier work is that in
\cite{Mitra10} there was an equator at $z=0$ with kinetic
helicity of opposite sign for $z<0$.
Consequently, also $\meanh_{\rm f}$ changes sign, allowing an efficient
exchange of magnetic helicity by the turbulence.
The present model is more similar to that of \cite{BD01},
except that there the kinetic helicity profile was flat in the
bulk of the dynamo interior and it dropped to zero only immediately
at the boundary of the domain, or gradually so in those cases where
a conducting non-turbulent halo was included.

The helicity in the halos, as noted above, will be suppressed by the low conductivity,
as we expect small-scale helicity transport away from the active central region.
A strong, rapidly achieved final mean-field would indicate that flux of small
scale helicity provides a clear escape from dynamical $\alpha$ quenching.  Finally,
a clear difference between halo simulations and simulations without a halo will
be evidence that the boundary conditions are generating artificial constraints.
If these differences are visible in the field itself (as opposed to the vector potential),
they are likely due to the reduced turbulent diffusion into the boundary.

Mean quantities are calculated from time series over a long stretch
of time where the relevant quantities are approximately stationary
in the statistical sense.
We use the time series further to calculate lower bounds on the error bars as the maximum
departure between these averages and the averages obtained from any of the
three equally long subsections of the full time series.

\begin{figure}[t!]\begin{center}
\includegraphics[width=\columnwidth]{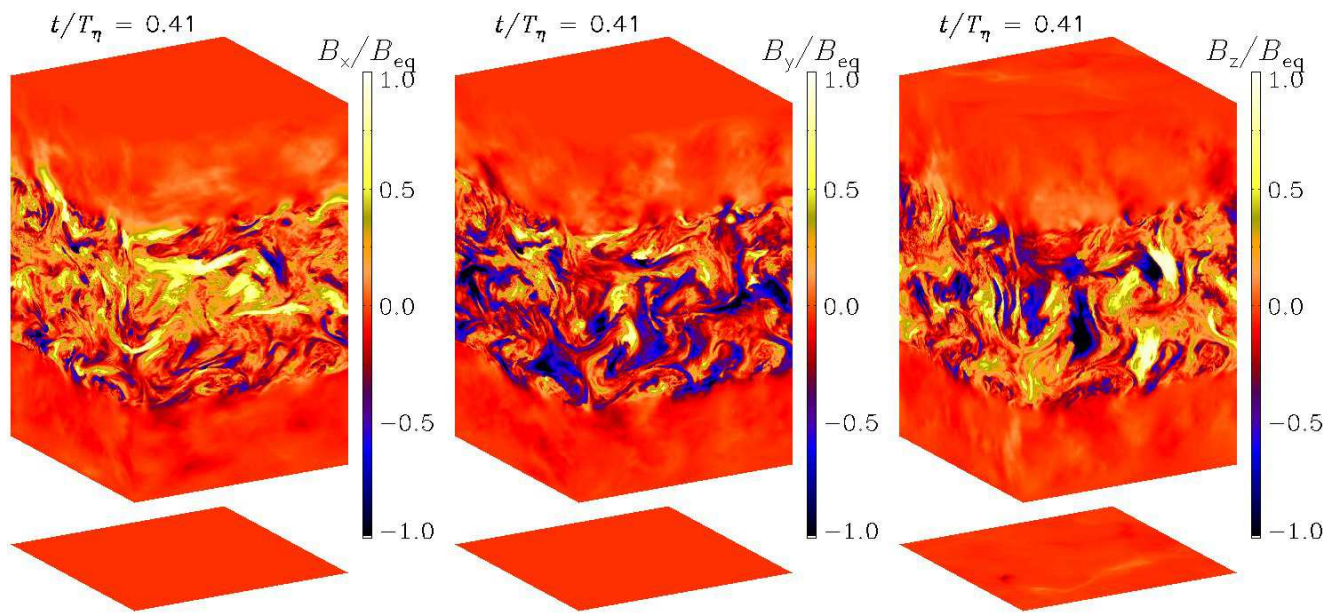}
\end{center}\caption[]{
Visualizations of each of the three components of $\BB$
on the periphery of the computational domain, for a run
with $\Rm=1300$ using $256\times256\times512$ mesh points.
The lower planes show the field at the lower boundary,
where it is quite weak.
Note the presence of mild overshoot into the upper and lower halos,
where $\eta_{\rm H}=250\eta_0$.
In the bulk of the domain the $x$ and $y$ components show a large-scale
field with variation in the $z$ direction, while $B_z$ does not show
a systematic mean field.
}\label{B}\end{figure}

\section{Results of simulations}
\label{analysis}

In the statistically steady state, the magnetic field shows a large-scale
magnetic field that varies in the $z$ direction; see \Fig{B}.
It is therefore meaningful to describe the dynamics of this large-scale
field by using horizontal averages as noted in \Sec{dyn}.  We will
use angular brackets and capitals to refer to volume averages
over the volume $V=L_x L_y L_z'$, where $L_z'=z_2-z_1$ 
and $-z_1=z_2=2k_1^{-1}$.
This will
mark the boundaries of a smaller domain well within the dynamo region.
Thus, we write
\EQ
H(t)=\langle h\rangle_V V
=\frac{1}{L_x L_y L_z'}\int\int\int h(x,y,z,t) \,\dd x\,\dd y\,\dd z
=\frac {1}{L_z'} \int \hhM(z,t)\,\dd z,
\EN
where $h$ and $H$ could stand for $H_{\rm m}$ and $h_{\rm m}$,
or for $H_{\rm f}$ and $h_{\rm f}$, for example.
We note, however, that these quantities may be gauge-dependent.
We also define the magnetic energy of the mean field as
$M_{\rm m}=\bra{\meanBB^2/2}V$.

\subsection{Small-scale helicity flux}

\begin{figure}[t!]\begin{center}
\includegraphics[width=\columnwidth]{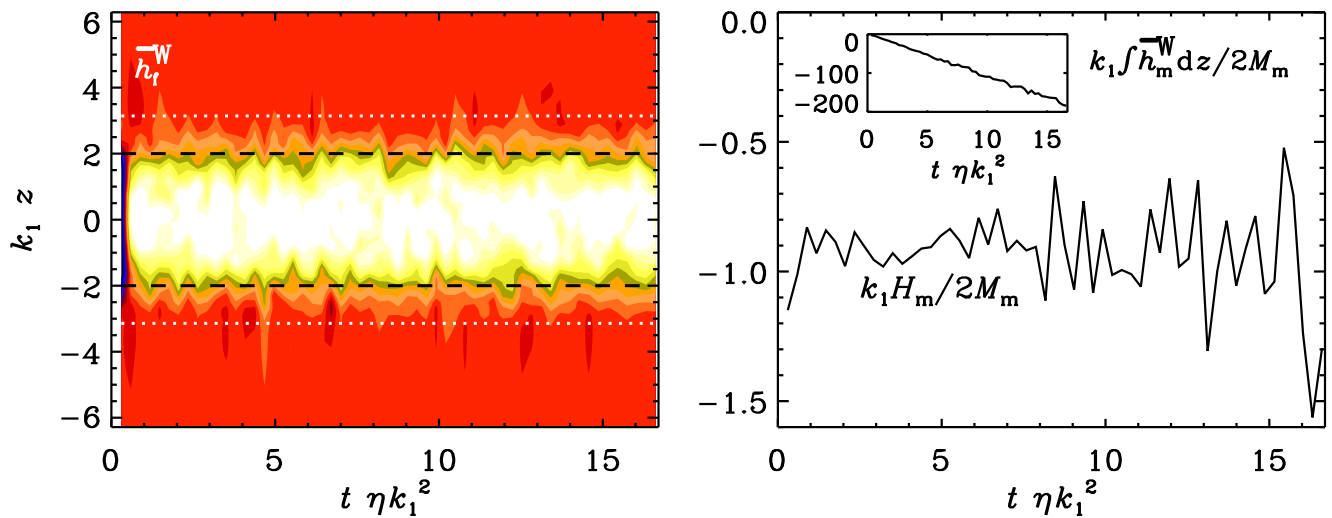}
\end{center}\caption[]{
Magnetic helicity density of the small-scale
magnetic field in the Weyl gauge, $\meanh_{\rm f}^{\rm W}$,
as a function of $z$ and $t$ (left-hand panel), and the
gauge-independent magnetic helicity of the large-scale
field, $H_{\rm m}$ in units of $2M_{\rm m}/k_1$ (right-hand panel),
for Run~H3 with $\eta_{\rm H}=100\eta_0$.
For comparison, the Weyl-gauged magnetic helicity of the large-scale
field is shown in the inset.
The data are averaged over regular time intervals of about 0.3 diffusive
times, which also explains the absence of data at $t=0$.
}\label{pflux_butter}\end{figure}

We define the magnetic helicity densities for the mean and fluctuating fields as
\EQ
\meanh_{\rm m}^{\rm W}=\meanAA\cdot\meanBB,\quad
\meanh_{\rm f}^{\rm W}=\overline{\aaaa\cdot\bb}.
\EN
The superscript W indicates that we are working in the Weyl gauge modulo
possible influences of the boundary conditions that have been mitigated through
the use of halos.
It turns out that $\meanh_{\rm m}^{\rm W}$ has a systematic variation in time
while $\meanh_{\rm f}^{\rm W}$ does not; see \Fig{pflux_butter}.
It makes therefore
sense to average the evolution equation for $\meanh_{\rm f}^{\rm W}$
in time, so we have \citep{Mitra10}
\EQ
\bbra{\partial\meanh_{\rm f}^{\rm W}\over\partial t}_T=0=
-2\bra{\meanEMF\cdot\meanBB}_T-2\eta\bra{\jj\cdot\bb}_T
-\bra{\nab\cdot\meanFFFF_{\rm f}^{\rm W}}_T,
\label{dhfdt}
\EN
where subscripts indicate time averaging over the interval $T$.
In the Weyl gauge the magnetic helicity flux of the small-scale field
is given by $\meanFFFF_{\rm f}^{\rm W}=\overline{\ee\times\aaa}$,
where $\ee=\EE-\meanEE$ is the electric field for the fluctuating quantities.
Given that the first two terms on the rhs of \Eq{dhfdt} are
gauge-invariant, $\bra{\nab\cdot\meanFFFF_{\rm f}}$ must also
be gauge-invariant, so we can drop the superscript W and
note that, in the particular case at hand, we have
$\bra{\nab\cdot\meanFFFF_{\rm f}^{\rm W}}_T
=\bra{\nab\cdot\meanFFFF_{\rm f}}_T$.
We emphasize that $\bra{\nab\cdot\meanFFFF_{\rm f}}_T$
is still a function of $z$.

\subsection{Large-scale helicity flux}

In order to assess the full magnetic helicity budget, we also
need to take the magnetic helicity of the mean field into account.
Since $\meanh_{\rm m}^{\rm W}$ is time-dependent, it is not possible
to invoke a similar argument as for $\meanh_{\rm f}^{\rm W}$.
We are therefore forced to abandon a detailed analysis of
the $z$ dependence of the magnetic helicity budget and restrict
ourselves to the analysis of the volume-integrated magnetic helicity,
$H_{\rm m}$, and its corresponding flux divergence, $Q_{\rm m}$,
using the gauge-invariant prescription of \cite{BD01}
for the volume in $z_1\leq z\leq z_2$,
\EQ
H_{\rm m}(t)=\int_{z_1}^{z_2}\meanh_{\rm m}^{\rm W}(z,t)\,\dd z
+\meanAA(z_1,t)\times\meanAA(z_2,t),
\EN
and
\EQ
Q_{\rm m}(t)=-\left[\meanEE(z_1,t)+\meanEE(z_2,t)\right]
\cdot\int_{z_1}^{z_2}\meanBB(z,t)\,\dd z, \label{Qdef}
\EN
where $\meanEE=\eta\meanJJ-\meanEMF$ is the mean
electric field expressed in terms of horizontally
averaged Ohm's law.
Note that $H_{\rm m}$ and $Q_{\rm m}$ obey the
evolution equation
\EQ
{\dd H_{\rm m}\over\dd t}=2\int_{z_1}^{z_2}\meanEMF\cdot\meanBB\,\dd z
-2\eta\int_{z_1}^{z_2}\meanJJ\cdot\meanBB\,\dd z-Q_{\rm m}.
\label{dhmdt}
\EN
It turns out that, unlike $\meanh_{\rm m}^{\rm W}$ and its
volume-integral, $H_{\rm m}$ is statistically steady
(see \Fig{pflux_butter}),
so we may now also average \Eq{dhmdt} over time.

\subsection{Magnetic helicity budgets}

In \Tab{pflux}, we summarize the helicity budgets, namely
the six terms on the rhs of \Eqs{dhfdt}{dhmdt}, of which
the $2\bra{\meanEMF\cdot\meanBB}$ term occurs twice.
We have used here the more descriptive symbol
$\bra{\nab\cdot\meanFFFF_{\rm m}}_{VT}\equiv Q_{\rm m}/(z_2-z_1)$
for the flux divergence of the helicity of the mean field.
In order to simplify the notation, we drop from now on the
subscripts $VT$ and define angular brackets without subscripts
as combined averages over a long enough time span and over the
volume $V$ in $z_1\leq z\leq z_2$, where
again, $-z_1=z_2=2k_1^{-1}$.

\begin{table}[b!]\caption{
Summary of the volume and time averaged terms on the rhs of
\Eqs{dhfdt}{dhmdt}, normalized by $\etatz\Beq^2$,
while $\bra{\meanBB^2}$ is normalized by $\Beq^2$.
Runs H1--6 refer to systems with a poorly conducting halo
with $\eta_{\rm H}/\eta_0\approx\Rm$,
while in Runs~VF1--6 refer to systems without a halo and a
vertical field boundary condition at $|k_1z|=\pi$.
The data for Run~H6 are given for completeness, but it has
not run long enough to have satisfactory statistics.
}\vspace{12pt}\centerline{\begin{tabular}{rrcccccc}
Run & $\Rm$ & $\bra{\meanBB^2}$ &
$2\bra{\meanEMF\cdot\meanBB}$ &
$2\eta\bra{\meanJJ\cdot\meanBB}$ &
$2\eta\bra{\overline{\jj\cdot\bb}}$ &
$\bra{\nab\cdot\meanFFFF_{\rm m}}$ &
$\bra{\nab\cdot\meanFFFF_{\rm f}}$ \\
\hline
H1 &  20&$0.56$&$-0.423\pm0.003$&$-0.068\pm0.000$&$ 0.408\pm0.002$&$-0.360\pm0.006$&$ 0.018\pm0.013$\\
H2 &  50&$0.33$&$-0.208\pm0.003$&$-0.018\pm0.000$&$ 0.190\pm0.001$&$-0.192\pm0.005$&$ 0.012\pm0.004$\\
H3 & 140&$0.15$&$-0.086\pm0.003$&$-0.003\pm0.000$&$ 0.078\pm0.001$&$-0.079\pm0.005$&$ 0.005\pm0.001$\\
H4 & 270&$0.12$&$-0.047\pm0.002$&$-0.001\pm0.000$&$ 0.041\pm0.000$&$-0.046\pm0.001$&$ 0.003\pm0.000$\\
H5 & 520&$0.08$&$-0.024\pm0.000$&$-0.000\pm0.000$&$ 0.020\pm0.000$&$-0.024\pm0.001$&$ 0.002\pm0.000$\\
H6 &1280&$0.08$&$-0.029\pm0.023$&$-0.000\pm0.000$&$ 0.009\pm0.000$&$-0.007\pm0.007$&$-0.007\pm0.004$\\
\hline
VF1 & 10&$0.62$&$-0.823\pm0.011$&$-0.163\pm0.002$&$ 0.822\pm0.005$&$-0.669\pm0.005$&$-0.000\pm0.012$\\
VF2 & 20&$0.43$&$-0.434\pm0.004$&$-0.051\pm0.002$&$ 0.436\pm0.003$&$-0.400\pm0.005$&$ 0.002\pm0.026$\\
VF3 & 50&$0.32$&$-0.250\pm0.013$&$-0.019\pm0.001$&$ 0.247\pm0.002$&$-0.224\pm0.009$&$ 0.006\pm0.012$\\
VF4 &120&$0.28$&$-0.138\pm0.009$&$-0.007\pm0.000$&$ 0.134\pm0.001$&$-0.143\pm0.004$&$ 0.001\pm0.008$\\
VF5 &220&$0.25$&$-0.091\pm0.002$&$-0.003\pm0.000$&$ 0.082\pm0.001$&$-0.086\pm0.002$&$ 0.004\pm0.003$\\
VF6 &400&$0.15$&$-0.053\pm0.002$&$-0.001\pm0.000$&$ 0.046\pm0.000$&$-0.052\pm0.001$&$ 0.005\pm0.002$\\
\label{pflux}\end{tabular}}\end{table}

The results given in \Tab{pflux} show that $2\bra{\meanEMF\cdot\meanBB}$
is balanced essentially by $\bra{\nab\cdot\meanFFFF_{\rm m}}$, because
$\eta\bra{\JJM \cdot\BBM}$ is small.
On the other hand, for the magnetic Reynolds numbers
considered here ($\Rm\la500$), the $2\eta\bra{\jj\cdot\bb}$
term is still quite large, and contributes mainly to balancing
$-2\bra{\meanEMF\cdot\meanBB}$ in the magnetic helicity balance for the
fluctuating field.
The other (smaller) contribution comes from $\bra{\nab\cdot\meanFFFF_{\rm f}}$.
This result is quite similar to that of \cite{Mitra10} for the
case of a linearly varying kinetic helicity profile, where it was
found that, even though most of $\bra{\meanEMF\cdot\meanBB}$ is still
balanced by $\eta\bra{\overline{\jj\cdot\bb}}$, both
$\bra{\meanEMF\cdot\meanBB}$ and $\bra{\nab\cdot\meanFFFF_{\rm f}}$
vary little with $\Rm$ and must eventually dominate over
$\eta\bra{\overline{\jj\cdot\bb}}$ as $\eta$ decreases with increasing $\Rm$.
This was estimated to happen at $\Rm=10^3...10^4$.
In the model presented here, this is not so obvious, because
$\bra{\meanEMF\cdot\meanBB}$ shows still a rapid decline with increasing $\Rm$.
This may be a consequence of the fact that also $\meanBB^2$ declines still
quite rapidly with increasing $\Rm$, which indicates that the quenching
is $\Rm$-dependent, at least for the values of $\Rm$ considered here.

Note that the final field strength for systems without halos tends to be
higher than for systems with halos: turbulent transport of the mean field
out of the active region plays an important role.
This implies that the turbulent
flux of magnetic helicity from small-scale fields has a weaker effect
on the final strength of the mean field than
the turbulent flux of the mean field itself.

Note also that the total helicities $H_{\rm m}$ and $H_{\rm f}$ show little
difference in the two setups, suggesting that the artificially imposed $h=0$ constraint
on the boundary is not generating spurious results.

\begin{figure}[t!]\begin{center}
\includegraphics[width=\columnwidth]{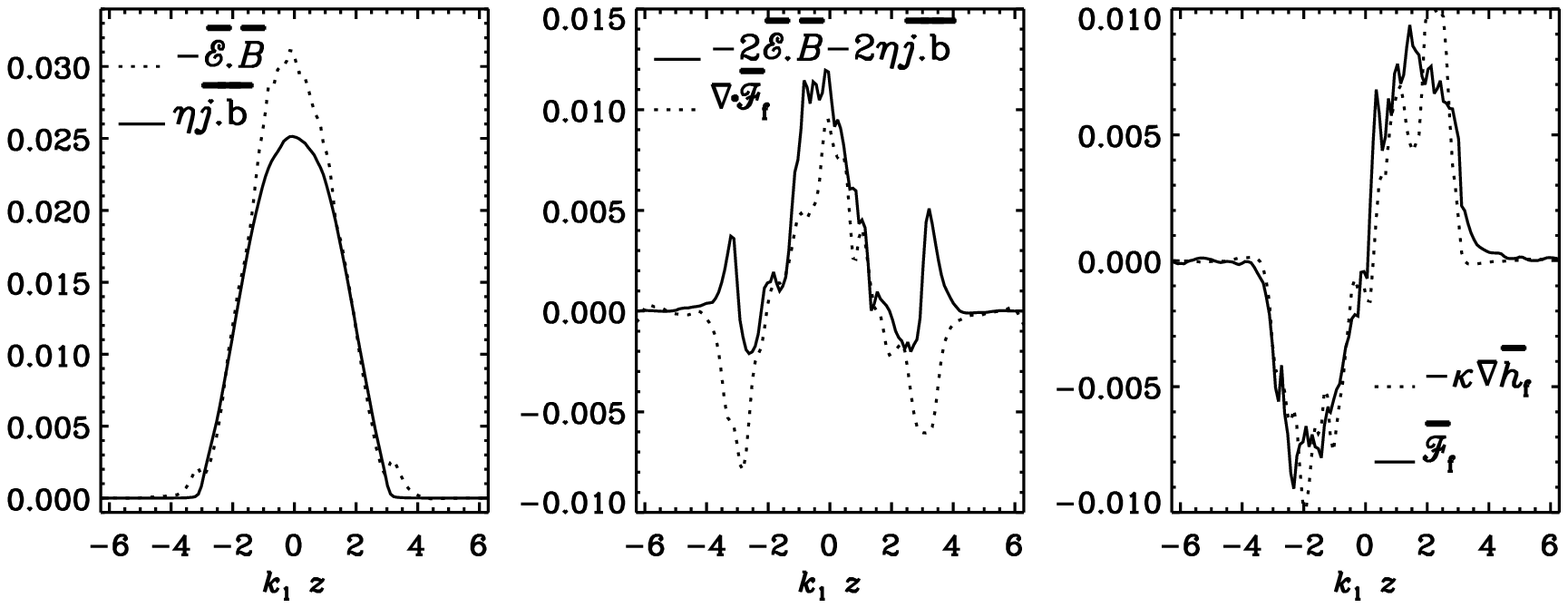}
\end{center}\caption[]{
Time-averaged terms on the right-hand side of \Eq{dhfdt} for Run~H4,
$\bra{\meanEMF\cdot\meanBB}_T$ and $\eta\bra{\jj\cdot\bb}_T$ (left panel),
the difference between these terms compared with the magnetic helicity flux
divergence of small-scale fields $\bra{\nab\cdot\meanFFFF_{\rm f}^{\rm W}}_T$
(middle panel), and the flux itself compared with the Fickian diffusion
ansatz (right-hand panel).
The fluxes are in given in units of $\etatz\Beq^2$ and the
flux divergence is given in units of $k_1\etatz\Beq^2$.
}\label{pflux_profile}\end{figure}
 
 \subsection{Magnetic helicity fluxes}
 
In \Tab{pflux2} we collect results for the magnetic helicity flux divergence.
The profile of the flux of magnetic helicity from the small-scale magnetic
field, $\meanFFFF_{\rm f}$, is reasonable well described by a Fickian
diffusion ansatz.
In \Fig{pflux_profile} we show the profiles of
$\bra{\meanEMF\cdot\meanBB}$ and $\eta\bra{\meanJJ\cdot\meanBB}$, compare the residual 
$2\bra{\meanEMF\cdot\meanBB}-2\eta\bra{\meanJJ\cdot\meanBB}$
with the divergence of the magnetic helicity flux, and finally
compare the flux $\meanFFFF_{\rm f}=\overline{\ee\times\aaaa}$
with that obtained from the diffusion approximation,
$-\kappa_{\rm f}\nab\meanh_{\rm f}$.

\begin{table}[b!]\caption{
Normalized values of magnetic helicity, current helicity,
and magnetic helicity flux divergence both for small-scale
and large-scale magnetic fields.
}\vspace{12pt}\centerline{\begin{tabular}{rrcccccc}
Run & $\Rm$ & 
$k_1H_{\rm m}/2M_{\rm m}$ &
$C_{\rm m}/k_1^2H_{\rm m}$ &
$Q_{\rm m}/\etatz k_1^2H_{\rm m}$ &
$k_1H_{\rm f}/2M_{\rm m}$ &
$C_{\rm f}/k_{\rm f}^2H_{\rm f}$ &
$Q_{\rm f}/\etatz k_1^2H_{\rm f}$ \\
\hline
H1&  20&$-0.94\pm 0.02$&$ 0.54\pm 0.02$&$ 0.69\pm 0.01$&$ 0.15\pm 0.00$&$ 2.05\pm 0.04$&$ 0.21\pm 0.15$\\
H2&  50&$-0.89\pm 0.01$&$ 0.56\pm 0.01$&$ 0.68\pm 0.02$&$ 0.21\pm 0.01$&$ 2.50\pm 0.12$&$ 0.18\pm 0.07$\\
H3& 140&$-0.93\pm 0.06$&$ 0.51\pm 0.01$&$ 0.55\pm 0.03$&$ 0.34\pm 0.00$&$ 3.02\pm 0.06$&$ 0.08\pm 0.02$\\
H4& 270&$-0.97\pm 0.02$&$ 0.50\pm 0.01$&$ 0.41\pm 0.01$&$ 0.38\pm 0.00$&$ 4.29\pm 0.07$&$ 0.08\pm 0.01$\\
H5& 520&$-0.90\pm 0.02$&$ 0.53\pm 0.01$&$ 0.34\pm 0.01$&$ 0.46\pm 0.00$&$ 4.96\pm 0.09$&$ 0.07\pm 0.01$\\
H6&1280&$-1.30\pm 0.01$&$ 0.36\pm 0.02$&$ 0.08\pm 0.07$&$ 0.36\pm 0.04$&$ 6.68\pm 1.40$&$-0.15\pm 0.18$\\
\hline
VF1 & 10&$-2.47\pm 0.45$&$ 0.20\pm 0.03$&$ 0.45\pm 0.08$&$ 0.11\pm 0.02$&$ 2.67\pm 0.52$&$-0.21\pm 0.36$\\
VF2 & 20&$-2.40\pm 0.43$&$ 0.21\pm 0.03$&$ 0.40\pm 0.08$&$ 0.16\pm 0.02$&$ 2.69\pm 0.31$&$ 0.02\pm 0.40$\\
VF3 & 50&$-2.26\pm 0.37$&$ 0.22\pm 0.03$&$ 0.31\pm 0.07$&$ 0.24\pm 0.03$&$ 2.68\pm 0.26$&$ 0.06\pm 0.14$\\
VF4 &120&$-1.82\pm 0.28$&$ 0.27\pm 0.04$&$ 0.29\pm 0.04$&$ 0.28\pm 0.01$&$ 3.49\pm 0.07$&$ 0.02\pm 0.10$\\
VF5 &220&$-1.84\pm 0.32$&$ 0.26\pm 0.05$&$ 0.19\pm 0.04$&$ 0.32\pm 0.00$&$ 3.83\pm 0.07$&$ 0.06\pm 0.04$\\
VF6 &400&$-1.49\pm 0.23$&$ 0.32\pm 0.04$&$ 0.23\pm 0.04$&$ 0.43\pm 0.01$&$ 4.76\pm 0.06$&$ 0.07\pm 0.04$\\
\label{pflux2}\end{tabular}}\end{table}

There are several additional points to be noted about the simulation results.
Firstly, based on earlier results for triply-periodic domains one expects
that $H_{\rm m}$ and $H_{\rm f}$ have the opposite sign, which is indeed
always the case.
Furthermore, we expect that the current helicity
of the mean fields, $C_{\rm m}\equiv \langle \JJM \cdot \BBM\rangle$,
and $H_{\rm m}$ have the same sign and that $C_{\rm m}/H_{\rm m}\approx k_1^2$.
This is indeed the case, except that the simulations give about half
or less than the naively expected value for $C_{\rm m}/H_{\rm m}$.
This indicates that the large-scale magnetic field is not fully helical,
a potential reason for the modest mean-field saturation strength
even in the presence of a magnetic helicity flux of small-scale fields.
Likewise, one expects that $C_{\rm f}$ and $H_{\rm f}$ have again the same sign.
Again, this is borne out by the simulations, but the ratio
$C_{\rm f}/k_{\rm f}^2H_{\rm f}$ is typically 3--5 times larger
than the expected value of unity.
This may well be a consequence of the presence of a finite
flux divergence of magnetic helicity of small-scale fields.

Finally, we find that the sign of the flux divergence of magnetic helicity
density of fluctuating and mean magnetic fields has the same sign as the
respective magnetic helicities themselves.
This is generally the case and is well motivated by the Fickian
diffusion ansatz.
Given that that we find $\kappa_{\rm f}\approx0.3\etatz$, we should
expect that $Q_{\rm f}/\etatz k_1^2H_{\rm f}$ is also about 0.3, but
the real value is only 0.1.
On the other hand $Q_{\rm m}/\etatz k_1^2H_{\rm m}$ varies between
0.2 and 0.7, but tends to decrease with increasing values of $\Rm$, although
it remains above $Q_{\rm f} /\etatz k_1^2 H_{\rm f}$ and may be approaching
it from above.  
As long as the transport of large-scale helicity has a larger
transport coefficient than that of the small-scale helicity,
we expect that the small-scale helicity transport will not
result in larger helical mean field strengths even though it
allows for a stronger post-kinematic $\alpha$ effect \citep{BS05}.

The complication that $\eta \langle \JJ \cdot \BB \rangle$
is not small even for the largest $\Rm$ should not be forgotten.
The total helicity being forced, and eventually even the halo
`buffer' zones will transmit information to the boundary.
It is therefore not clear how well a Fickian diffusion ansatz is
justified for the helicity of the mean magnetic field.

\section{Connection with mean-field models}
\label{meanfield}

\begin{figure}[t!]\begin{center}
\includegraphics[width=\columnwidth]{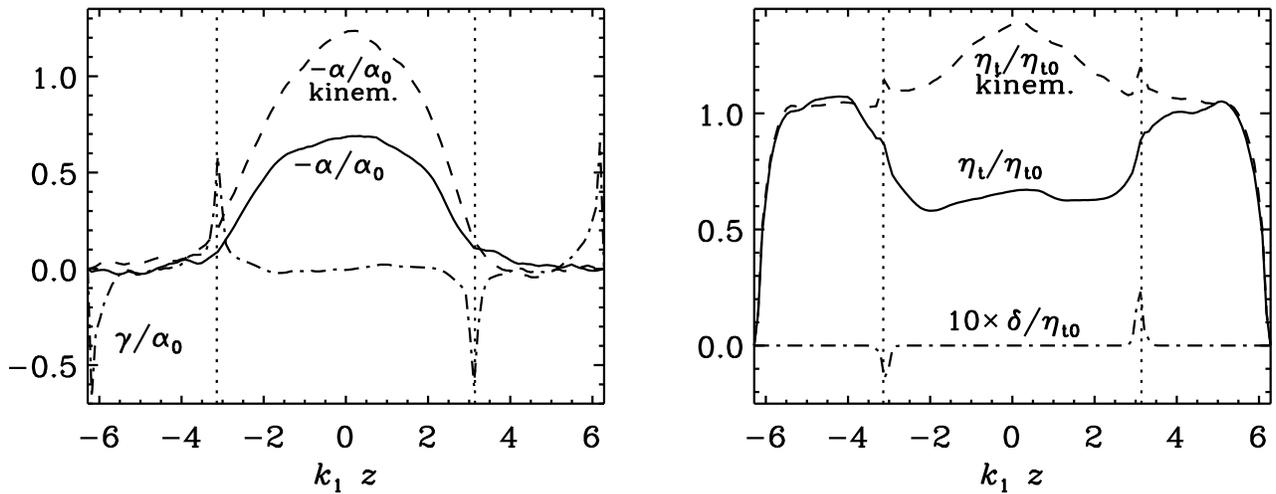}
\end{center}\caption[]{
Profiles of $\alpha$ and $\etat$ for Run~H4
for the saturated case (solid lines) and the
kinematic case (dashed lines),
obtained using the test-field method.
In the left and right panels, we also show $\gamma$
and $\delta$, respectively.
}\label{ptest}\end{figure}

In order to perform mean-field simulations, we need to include all the relevant
turbulent transport coefficients.
A robust tool for extracting these coefficients from simulation
is the test-field method of \cite{Sch05,Sch07};
for applications to time-dependence turbulence see \cite{BRS08,BRRK08}.
We apply this technique both to the kinematic and to the nonlinear stage
using the so-called quasi-kinematic test-field method \cite{BRRS08};
for a justification of it see \cite{RB10}.
In \Fig{ptest} we show not only the values of
$\alpha$ and $\etat$ as determined by the test-field method, but also
the $\gamma$ and $\delta$ effects in the more general expression
\EQ
\EMFM =\alpha \BBM+\gamma\zzz\times\BBM
-\etat \JJM+\delta\zzz\times\JJM.
\EN
As expected, the latter are negligible.
Interestingly, we see evidence of quenching of $\etat$ in the active region,
even though the mean-field is well below equipartition ($\langle \BBM^2\rangle
=0.1 \Beq^2$).  Further, we see an approximately 2-fold reduction
both in $\alpha$ and in $\etat$.

\begin{figure}[t!]\begin{center}
\includegraphics[width=\columnwidth]{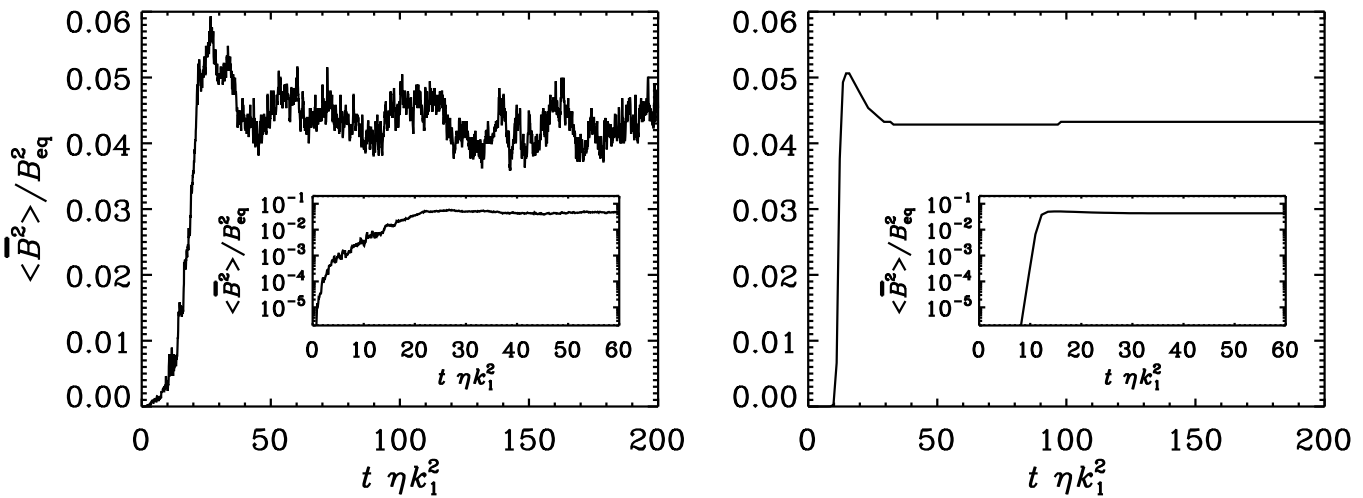}
\end{center}\caption[]{
Comparison between the saturation behavior in the simulation (left) and
the mean-field model (right) for the same parameters:
$\Rm/3=100$, $C_\alpha=3$.
Magnetic helicity flux and its fit in terms of a Fickian diffusion law,
for Run~H4.
}\label{psat}\end{figure}

In \Fig{psat} we compare the evolution of $\bra{\meanBB^2}/B_{\rm eq}^2$
for the direct simulation with the solution of the corresponding mean-field model.
There is excellent agreement in the final saturation level, and in
both cases the amplitude overshoots slightly before settling at
a somewhat lower value, but the kinematic growth rate is much
faster in the mean-field model than in the simulation.
This discrepancy is not yet well understood and should be reconsidered
in future work.
Perhaps significantly, the rise time of the mean field is rapid
even in terms of the turbulent turnover time.
This means that memory effects become important \citep{HB09},
and that the actual growth rate would be reduced compared with
that obtained from simple estimates.
The overshoot may simply be an artifact of the finite time it takes the system
to convert mean fields into correlated small-scale fields,
which is not included in the mean-field model.

\section{Conclusions}
\label{conclusions}

Confirming earlier work of \cite{Mitra10}, we have demonstrated the
existence of a diffusive flux $\meanFFFF_{\rm f}$ of mean magnetic
helicity of the small-scale field.
In the present case, however, the Weyl-gauged magnetic helicity
of the large-scale field never reaches a steady state.
Nevertheless, the magnetic helicity density of the small-scale magnetic
field is found to be statistically steady, so the corresponding
magnetic helicity flux must be gauge-independent \citep{Mitra10}.
This supports the validity of using the small-scale magnetic helicity as
a meaningful proxy for the small-scale current helicity,
and hence the magnetic correction to the $\alpha$ effect.

Understanding the transport of magnetic helicity of the large-scale
field, $\meanFFFF_{\rm m}$, would be useful for creating analytic
post-kinematic models.
Although we have not converged on a formula for this flux,
it is certainly finite and apparently $\Rm$ dependent.
It is not yet clear whether this flux will converge to a diffusive one for large $\Rm$.
Our mean-field
simulations reproduce the final field strength well, reinforcing the conclusion that
post-kinematic dynamical $\alpha$ quenching can be used as
part of a mean-field simulation.

The preliminary evidence on the use of small-scale helicity fluxes to escape the small
predicted post-kinematic mean fields is negative: the observed flux of large-scale helicity,
while poorly modeled, is larger than the flux of the small-scale helicity.
If this holds
for larger $\Rm$, it would have the unfortunate result of closing escape holes
from $\alpha$ quenching opened by $\meanFFFF_{\rm f}$, but would also imply that
dynamo systems with more realistic profiles than simple homogeneity
will reach $\Rm$ independent behavior for high but currently nearly
numerically achievable $\Rm$.
It is likely that conclusive evidence for or against $\Rm$-dependent quenching
requires values of $\Rm$ in the range between $10^3$ and $10^4$ \citep{BCC09,Mitra10}.

\section*{Acknowledgements}

We thank Petri K\"apyl\"a for detailed suggestions that have
helped to improve our paper.
We acknowledge the allocation of computing resources provided by the
Swedish National Allocations Committee at the Center for
Parallel Computers at the Royal Institute of Technology in
Stockholm and the National Supercomputer Centers in Link\"oping
as well as the Norwegian National Allocations Committee at the
Bergen Center for Computational Science.
This work was supported in part by
the European Research Council under the AstroDyn Research Project No.\ 227952
and the Swedish Research Council Grant No.\ 621-2007-4064.

\vfill\bigskip\noindent\tiny\begin{verbatim}
$Header: /var/cvs/brandenb/tex/hubbard/HelicityFluxShort/paper.tex,v 1.76 2010-04-26 18:04:44 brandenb Exp $
\end{verbatim}

\end{document}